\documentclass[10pt]{article}
\usepackage{graphicx}
\title{Mach's Principle and Cosmology Term}

\author{V.M.Lipunov}

\date{}
\begin{document}
\maketitle

\begin{abstract}

Recent discovery of the rate of expansion of the Universe (Reiss A.G. et al.,
1998; Perlimutter S. et al., 1999) is connected with the vacuum energy in space
which, in turn, can be considered (Gliner E.B., 1965) as the cosmological
constant in Einstein's equation. In this article the cosmological term
predicted by Einstein is interpreted in terms of World's mass required by
Mach's principle. The Mach's principle itself is formulated as follows: inertia
masses of bodies, including those of fundamental particles, are determined by the
value of the cosmological term.

\end{abstract}

Discovery of the rate of expansion of the Universe due to Ia supernovae
observations (Reiss A.G. et al., 1998; Perlimutter S. et al., 1999), analysis of
relict radiation fluctuations and studies of large-scale matter distribution
(Wang L. et al., 2000) have proved existence of the cosmological term that
Einstein added to the equations for his theory.

The cosmological term is believed to have been introduced by Einstein (1917)
for a static Universe to be created.

$$
G_{\mu\nu}-\lambda g_{\mu\nu} = -\kappa \left(T_{\mu\nu} -\frac12 g_{\mu\nu} T \right)
$$

The cosmological term thus postulated is equivalent to global antigravitation which 
is able to balance self-gravitation of the matter for the Universe to be static 
(under certain conditions). The most important Einstein's task and the mainspring 
of his works, however, was an effort to match the
equations of General Theory of Relativity (GTR) with Mach's principle. Moreover, 
the appearance of the relativistic theory of gravitation was due to, at least from the 
ideological point of view, Mach's principle itself. 
Contemporary physics, however, following latest Einstein's works doesn't consider Mach's
principle as an essential requirement. The cosmological term, subsequent to Gliner (1965), 
is interpreted as vacuum energy and vacuum itself is assumed to be medium with certain 
energy density (epsilon) which satisfies 
Lorentz invariant equation of state:

\begin{equation} 
p = -\epsilon
\label{eq_state}
\end{equation}

The fact that vacuum possesses energy follows from zero oscillations discovered owing to 
Lamb shift measurements (see a nice review by Chernin (2001)). 
In this paper we are going to make an attempt to apply Mach's principle to contemporary conditions.

\section{Mach's principle as a main principle of General Theory of Physics}

In his paper \cite{einstein16}, after having discussed Mach's principle, Albert Einstein gave logical 
foundations of the General Theory of Relativity. It's 
worthwhile to remind ourselves that Mach's principle relates
inertia of a body to other mass existence.
The principle can be briefly formulated as follows: there is no inertia in a void Universe. 
Both Newtonian mechanics and special theory of relativity suffer from one fundamental disadvantage,
as Einstein explains using a simple example. Let, in an absolutely void Universe, 
two identical liquid bodies rotate around the common axis, the centers of the bodies being 
located on the axis. Two observers measure the shapes of the bodies and come to alternative results. 
One of them reports, since a body is at rest relative to the void space $O_1$, it is of a spherical 
shape; another informs, since the body rotates relative to the void space $O_2$, it is elongated 
along the equator. Who is right - one may decide, from a theoretical and cognitive point of view, 
only if a cause that preceded the consequence might be observable in principle. It is obvious, 
as long as the spaces $O_1$ and $O_2$ are void, they can not be distinguished from each other. 
That means, the differences between the spaces (if any) can not be observed. 
Therefore, concludes Einstein, a space must be defined by some extra masses. 
As a matter of fact, this immediately implies that a theory of gravitation must be a geometrical one, 
that is, a theory in which the space-time continuum is defined and determined by a distributed matter.
It was a natural temptation, in view of relative rareness of our space, to move such "causative"
masses away to infinity. Einstein did that, he made an attempt to create a kind of cosmology 
(the space-time continuum) where masses were removed to infinity. This attempt, however, failed;
boundary conditions could not be specified at infinity and, besides, the Universe was no longer 
static (due to attraction of the aforementioned masses). It was the moment when a witty solution
came to his mind: if it is impossible to set a priory universal boundary conditions at infinity, 
one ought to get rid of the infinity itself, and the question will no longer be relevant. 
That is, one should create a static, closed world. Namely then Einstein arrived at necessity to 
introduce antigravitation, the Lambda-term, which fortunately did not influence the requirement 
according to which GTR equations must be covariant. It was a strange Universe: attraction of 
distributed homogeneous matter was counterbalanced by antigravitaton of the cosmological term but 
even a small pushing of the world would result in its contracting or expanding away. And this was 
inconceivable, in that case a universal feature of all bodies, their inertia mass, would be dependent 
on time.
	
It is de Sitter\cite{desitter} who saved the situation. He noticed that constructing static 
cosmology did not need any extra masses; what was really needed, was the Lambda-term in a "
void" space. De Suitter's cosmology of a "void world" is described by the following metrics:

$$ 
ds^2= S(r) dt^2 -r^2 d\Omega^2 - S^{-1}(r) dr^2
$$
$$
S(r)=\left(1-\left(\frac{r}{A_V}\right)^2\right)
$$

When discussing Mach's principle, de Sitter analyzed Einstein's failures as to introducing
"remote" and, later, "distributed" world's masses. He concluded that Mach's principle was to
be given up altogether (at least, in its original form). After all, the main requirements - static 
character and boundary condition specifications - were answered in an entirely void Universe. 
Later A.A.Fridman (1920 - 1924) developed the most thorough cosmological solution which involved 
both the Lambda-term and the matter. Einstein's cosmology and that of de Sitter are only particular
cases of this solution.

It might seem that Mach's principle was totally wrecked. In a void Universe there is a geometry, 
hence there is gravitation which implies inertia without matter!
	
\section{Vacuum energy as "world's" or "causative" mass}

Now, after having created quantum theory and appearance of Gliner's work\cite{gliner65}, 
we understand that introducing cosmological term into the left part of Einstein's equations is 
equivalent to introducing vacuum energy and appearing energy-momentum tensor of a void space
on the right of equation. Contemporarily speaking, de Sitter did not "devastated" the space-time 
continuum, on the contrary, he filled it out homogeneously with the "world's mass" being vacuum energy.
To put it another way, Mach's principle demanded the presence of the Lambda-term or vacuum energy. 
With this viewpoint in mind, Mach's principle can be formulated, in terms of modern conceptions, 
as follows:

\emph{
Masses of fundamental particles are quantitatively caused by the value of the Lambda-term. 
If the Lambda-term were 0, masses would be 0, too. 
}

This wording seems to be too much audacious basically for two reasons. 

\begin{enumerate}
\item{Vacuum state is of Lorentz-invariant character. Energy of the vacuum does not depend on the 
motion of a body and, in addition, does not impede its motion.}
\item{A contemporary quantum field theory predicts the vacuum energy $100$ times of that discovered 
in astronomical observations.}
\end{enumerate}
 
We do not consider these objections to be critical. Actually, they may be of temporary character.

First, it is the introduction of the Lambda-term that allows the space-time framework to be 
constructed and the motion itself to be observable. Thus, phenomenon of inertia, i.e. body's
willingness to keep on motion (changing coordinates) may be considered to result from the existence
of the space-time continuum.

Second, the grand unification theory is now far from its completion and, generally speaking, 
it can not so far provide an adequate value of the vacuum energy.

Therefore, according to a new Mach's principle, the cosmological term could not theoretically be 
equal to zero. Masses of all fundamental particles can be assumed to be an even power of the 
cosmological term when it aims at zero:

$$
M \propto \Lambda^{2n}\mbox{   , if   } \Lambda\to 0
$$

where $n$ is a natural number. At least, since Lambda may as well be below zero the first term 
in its expansion must be missing.

\section{ Contemporary cosmology with the cosmological term}

Discovery of the Lambda-term having a value of about $0.7$ (in terms of the critical density) 
has changed dramatically our ideas concerning global structure of the Universe (Chernin, 2001). 
The situation now may be described as follows (for details see the review by Chernin (2001)). 

The cosmology of our Universe, i.e. the Universe filled with matter and vacuum energy, 
is completely described by a general solution that was discovered by Fridman 
(do not confuse with the "Fridman model" which is a particular case of Fridman's solution). 
The world's dynamics, i.e. changes of the scale factor a with time, is given by the equation:

$$
\frac12 \dot a^2 = A_V^{-2}a^2+\frac{A_D+A_B}{a}+\frac12\frac{A_R^2}{a^2}-\frac{K}{2}
$$

Here, $A$-constants are Fridman's ones for vacuum, dark matter, baryons, and radiation,
correspondingly; $K$, a curvature sign: $1$, $0$, $-1$ for closed, plain, and open models.

\begin{figure}
\includegraphics{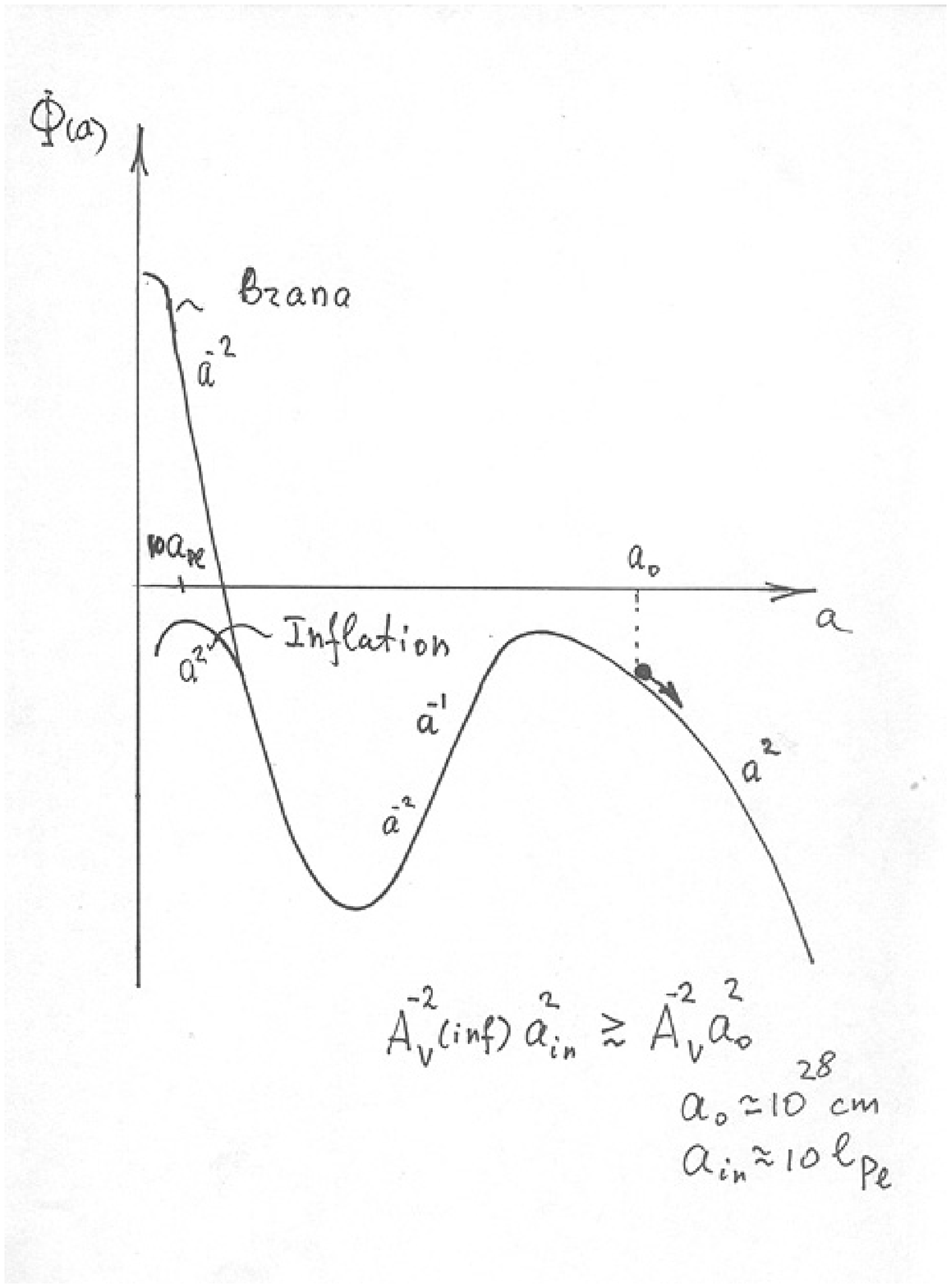}
\label{fig1}
\caption{}
\end{figure}

The left-hand part of the equation is actually a potential energy or an effective potential of 
the Universe (Fig.\ref{fig1}). In this Figure, the modern position of the Universe corresponds 
to its sliding down along the vacuum parabola. In itself, the cosmology can not answer 
the question how our Universe arose or why we witness its actual speeded-up expansion now. 
Having used outer (with respect to GTR) hypotheses allowed a so-called inflationary scenario to be
developed (Gliner, 1965; Gliner and Dymnikova, 1975; Starobinsky, 1979; Linde, 1981; Guth, 1981). 
In such a scenario the expansion is due to pushing aside due to the vacuum  energy. 
It is worthwhile mentioning, however, that the inflationary approach now, after the discovery of the
Lambda-term, lost a lot in its beauty. Actually, nowadays one needs to accept that energy of the
vacuum is not a constant value but rather a dynamical variable. Its original value which resulted in
changing space scale due to inflation by a hundred of orders, exceeds the value of the vacuum energy 
which was discovered by astronomers. 

The situation seems to be quite artificial and, per se, in conflict with the conceptions of GTR 
where the cosmological term may be only constant. A scenario with brane seems much more attractive 
(see, for example, Binetrue et al., hep-th/9910219, preprint)) where the cause of
inflation is a consequence of a brane in 5-fold space and in no way is related to the vacuum energy.

Actually, the potential involves a positive term growing into the past.

Although we know nothing on the nature of the "original momentum" we can still impose certain 
constraints on its absolute value. The Universe can overcome a secondary maximum being due to
self-attraction if

$$ 
A_{brane} > \frac{a_0 a_{in}}{A_V}
$$
Assuming an initial radius of the Universe $a_{in}$ to be about $10$ Planck scales yields

$$
A_{brane}> 10^{-32}
$$

The destiny of our Universe is likely to be a total recession of matter 
(if the value of the Lambda-term is going to be confirmed), the space-time framework turning 
into something very similar to de Sitter's world, static and void.

\section{Conclusion: Mach's principle and quantum theory}

In conclusion of this short paper it is worthwhile to mention one amazing circumstance: 
General Theory of Relativity which was developed only as a geometrical theory "managed to 
foresee" the quantum theory as well as one of its main consequences, namely, presence of energy 
in a void space. If we recall now that one of the basic demands of the theory was fulfillment of 
Mach's principle, i.e. conditionality of inertia by world's masses, the mentioned circumstance seems
to be rather startling. It may be related to some properties of the physical world yet undiscovered.
Naturally, our hypothesis on relation between inertia of bodies and the cosmological term needs 
further development and may be confirmed only in terms of the entire theory of quantum gravitation.

\section*{Acknowledgements}
I express my gratitude to A.I.Zakharov, A.P.Orlov, A.A.Starobinsky for their critical remarks and
useful discussions.

\end{document}